\documentclass{PoS}

\usepackage[utf8]{inputenc}
\usepackage{mathrsfs}
\usepackage{amsmath}
\usepackage{amssymb}
\usepackage{color}
\usepackage[caption=false]{subfig}

\newcommand{\nc}{\newcommand}

\newcommand{\Eqn}[1]{Eq.~(\ref{#1})}

\nc{\scm}{\scriptscriptstyle\mathrm}
\nc{\f}{\frac}
\nc{\be }{\begin{equation}}   \nc{\ee }{\end{equation}}
\nc{\bea}{\begin{eqnarray}}   \nc{\eea}{\end{eqnarray}}
\nc{\baa}{\begin{array}}      \nc{\eaa}{\end{array}}
\nc{\bit}{\begin{itemize}}    \nc{\eit}{\end{itemize}}
\nc{\ben}{\begin{enumerate}}  \nc{\een}{\end{enumerate}}
\nc{\bce}{\begin{center}}     \nc{\ece}{\end{center}}
\nc{\bfl}{\begin{flushright}} \nc{\efl}{\end{flushright}}
\nc{\btb}{\begin{tabular}}    \nc{\etb}{\end{tabular}}
\nc{\eps}{\varepsilon}
\nc{\vp}{\varphi}
\nc{\tvp}{\widetilde{\varphi}}
\nc{\D}{\mbox{$\not\!\!D$}}
\nc{\Db}{\mbox{${\raisebox{2mm}{\boldmath ${}^\leftarrow$}\hspace{-4mm} D}$}}
\nc{\Dfb}{\mbox{$\raisebox{2mm}{\boldmath ${}^\leftrightarrow$}\hspace{-4mm} D$}}
\nc{\vpj }{\mbox{${\vp^\dag i\,\,\raisebox{2mm}{\boldmath ${}^\leftrightarrow$}\hspace{-4mm} D_\mu\,\vp}$}}
\nc{\vpjt}{\mbox{${\vp^\dag i\,\,\raisebox{2mm}{\boldmath ${}^\leftrightarrow$}\hspace{-4mm} D_\mu^{\,I}\,\vp}$}}

\title{Leptonic CP violation in the charged sector\\
 and effective field theory approach}

\ShortTitle{Leptonic CP violation in the charged sector and EFT approach}

\author{\speaker{Giovanni Marco Pruna}
\\
        Paul Scherrer Institut,CH-5232 Villigen PSI, Switzerland\\
        E-mail: \email{Giovanni-Marco.Pruna@psi.ch}}

\abstract{
These proceedings introduce the main techniques and ideas for a systematic effective field theory analysis of CP violation in the charged lepton sector. This study is required because the Standard Model of particle physics predicts a very high degree of CP violation suppression in the lepton sector, and this implies that possible new physics effects can be parameterised in terms of new interactions among the Standard Model fields, in the framework of the so-called Standard Model Effective Field Theory. In analogy with previous investigations of charged lepton flavour violating observables, this document illustrates how the current limits on leptonic CP violation coming from the electric-dipole moment of the leptons can be recast into constraints on the effective coefficients defined at a given decoupling scale. Furthermore, important bounds acting on previously unconstrained parameters are discussed.}

\FullConference{The 15th International Conference on Flavor Physics \& CP Violation\\
		5-9 June 2017\\
		Prague, Czech Republic}

\begin{document}

\section{Introduction}
\label{sec:1}
\noindent
The Standard Model (SM) of particle physics fails to explain the origin of matter, the nature of neutrino oscillations, the observation of dark matter and dark energy satisfactorily, and it does not accommodate gravity. Consequently, it is accepted as a low-energy manifestation of an ultimate theory, defined at the Planck energy scale, which should incorporate solutions to these open problems.

With the assumption that these unexplained observables find their natural solution at some energy higher than the electroweak (EW) symmetry-breaking scale, the impact of such new physics (NP) can be well described by the SM effective field theory (SMEFT)~\cite{Buchmuller:1985jz,Grzadkowski:2010es,Henning:2014wua,Lehman:2014jma,Passarino:2016pzb,Brivio:2017vri,Celis:2017hod}. This theoretical framework consists of a parameterisation of the NP in terms of new interactions associated with non-renormalisable operators and is extremely efficient for the study of observables related to vanishing or suppressed SM predictions.

One class of such observables is related to CP violation in the lepton sector, which is completely absent at the tree level in the SM. Nevertheless, the observed baryon-antibaryon asymmetry in the Universe~\cite{Dine:2003ax} calls for enhanced CP violation with respect to the amount provided by the Cabibbo--Kobayashi--Maskawa (CKM) matrix of the SM~\cite{Canetti:2012zc}, and there is yet no experimental evidence for the elusive $\theta_{{\rm QCD}}$ term~\cite{Afach:2015sja}. Therefore, it is assumed that an extra contribution could come from beyond-the-SM (BSM) physics at higher energies. Such NP could also give rise to non-vanishing CP-violating contributions in the lepton sector~\cite{Branco:2011zb,Dinh:2017smk}.

The most relevant example of a CP-violating observable in the charged lepton sector at low energy is the electric-dipole moment (EDM), which is predicted to be very small in the SM~\cite{Pospelov:1991zt,Pospelov:2005pr,Engel:2013lsa} and so far has not been observed in precision experiments~\cite{Safronova:2017xyt}. This makes it especially suitable for analysis within a SMEFT framework.

In these proceedings, a systematic SMEFT analysis of lepton EDMs is performed up to the one-loop level with the inclusion of dimension-six operators, and the current experimental bounds are recast into constraints on the effective coefficients defined at higher energy scales.

In Section~\ref{sec:2}, the lepton sector of the SMEFT, with the inclusion of dimension-six operators, is presented together with a low-energy EFT valid at the lepton mass scale. Section~\ref{sec:3} describes the relation between leptonic EDM and effective coefficients and current experimental limits are listed. In Section~\ref{sec:4}, the connection between the effective coefficients defined above and below the EW scale and their mixing effects are explained. In Section~\ref{sec:5}, experimental bounds are interpreted in terms of limits on the effective coefficients at the EW scale and above. Finally, Section~\ref{sec:6} briefly discusses prospects for SMEFT analysis in connection with lepton EDM.

\section{Standard Model Effective Field Theory for Charged Leptons}
\label{sec:2}
\noindent
In this section, a set of SMEFT operators is formally introduced and discussed in the light of lepton EDM relevant phenomenological aspects.

First, BSM is assumed to generate non-standard interactions at some large energy scale, $\Lambda_{{\rm UV}}$. Exploiting the Appelquist--Carazzone theorem~\cite{Appelquist:1974tg}, these new interactions are parameterised by higher dimensional operators:
\begin{align}
\mathcal{L}_{\rm SMEFT} = \mathcal{L}_{\rm SM} + 
\frac{1}{\Lambda_{{\rm UV}}} \sum_i C^{(5)}_i\, Q^{(5)}_i + 
\frac{1}{\Lambda_{{\rm UV}}^2} \sum_i C^{(6)}_i\, Q^{(6)}_i +\left[\dots\right],
\label{Lag6}
\end{align}
where the complete set of operators up to dimension-seven are presented and discussed in~\cite{Buchmuller:1985jz,Grzadkowski:2010es,Lehman:2014jma}. The dimension-five operator studied in~\cite{Petcov:1976ff, Minkowski:1977sc} is severely constrained by the smallness of the neutrino mass scale, and operators with a dimension higher than six are scale-suppressed by construction. For the purpose of illustration, only the leptonic SMEFT operatorial classes listed in
Tables~\ref{tab:no4ferm} and~\ref{tab:4ferm} are considered
. In principle, more non-leptonic dimension-six operators could have an impact on leptonic EDM via quantum fluctuations~\cite{Alonso:2013hga}, but a complete treatment of leading and subleading effects is beyond the scope of this discussion and will be presented in a future publication~\cite{future}. The notation and conventions adopted in the following analysis are taken from~\cite{Pruna:2014asa}.

\begin{table}[!ht] 
\centering
\renewcommand{\arraystretch}{1.5}
\begin{tabular}{||c|c||c|c||c|c||} 
\hline \hline
\multicolumn{2}{||c||}{$\psi^2 X\vp$} & 
\multicolumn{2}{|c||}{$\psi^2\vp^2 D$} &
\multicolumn{2}{|c||}{$\psi^2\vp^3$}\\
\hline
$Q_{eW}$  & $(\bar l_p \sigma^{\mu\nu} e_r) \tau^I \vp W_{\mu\nu}^I$&  
$Q_{\vp l}^{(1)}$  &$(\vpj)(\bar l_p \gamma^{\,\mu} l_r)$ &
$Q_{e\vp}$           & $(\vp^\dag \vp)(\bar l_p e_r \vp)$\\
$Q_{eB}$  & $(\bar l_p \sigma^{\mu\nu} e_r) \vp B_{\mu\nu}$&   
$Q_{\vp l}^{(3)}$ & $(\vpjt)(\bar l_p \tau^I \gamma^{\,\mu} l_r)$&
 & \\
 & &    
$Q_{\vp e}$  & $(\vpj)(\bar e_p \gamma^{\,\mu} e_r)$&
 & \\
\hline \hline
\end{tabular}
\caption{Leptonic dimension-six operators consisting of fermions and bosons, according to~\cite{Grzadkowski:2010es}.\label{tab:no4ferm}}
\end{table}

\begin{table}[!ht]
\centering
\renewcommand{\arraystretch}{1.5}
\begin{tabular}{||c|c||c|c||}
\hline\hline
\multicolumn{2}{||c||}{$(\bar LL)(\bar LL)$} & 
\multicolumn{2}{|c||}{$(\bar RR)(\bar RR)$}\\
\hline
$Q_{ll}$    & $(\bar l_p \gamma_\mu l_r)(\bar l_s \gamma^{\,\mu} l_t)$ &
$Q_{ee}$    & $(\bar e_p \gamma_\mu e_r)(\bar e_s \gamma^{\,\mu} e_t)$ \\
$Q_{lq}^{(1)}$& $(\bar l_p \gamma_\mu l_r)(\bar q_s \gamma^{\,\mu} q_t)$&  
$Q_{eu}$     & $(\bar e_p \gamma_\mu e_r)(\bar u_s \gamma^{\,\mu} u_t)$\\
$Q_{lq}^{(3)}$  & $(\bar l_p \gamma_\mu \tau^I l_r)(\bar q_s \gamma^{\,\mu} \tau^I q_t)$&
$Q_{ed}$     &$(\bar e_p \gamma_\mu e_r)(\bar d_s\gamma^{\,\mu} d_t)$ \\
\hline\hline
\multicolumn{2}{||c||}{$(\bar LL)(\bar RR)$}&
\multicolumn{2}{|c||}{$(\bar LR)(\bar RL)$ and $(\bar LR)(\bar LR)$}\\
\hline
$Q_{le}$    & $(\bar l_p \gamma_\mu l_r)(\bar e_s \gamma^{\,\mu} e_t)$ &
$Q_{ledq}$   &  $(\bar l_p^j e_r)(\bar d_s q_t^j)$\\
$Q_{lu}$     & $(\bar l_p \gamma_\mu l_r)(\bar u_s \gamma^{\,\mu} u_t)$&
$Q_{lequ}^{(1)}$&$(\bar l_p^j e_r) \eps_{jk} (\bar q_s^k u_t)$ \\
$Q_{ld}$     & $(\bar l_p \gamma_\mu l_r)(\bar d_s \gamma^{\,\mu} d_t)$&
$Q_{lequ}^{(3)}$&$(\bar l_p^j \sigma_{\mu\nu} e_r) \eps_{jk} (\bar q_s^k \sigma^{\mu\nu} u_t)$ \\
$Q_{qe}$     & $(\bar q_p \gamma_\mu q_r)(\bar e_s \gamma^{\,\mu} e_t)$&
            &                                                   \\
\hline\hline
\end{tabular}
\caption{Leptonic dimension-six operators consisting of four fermions, according to~\cite{Grzadkowski:2010es}. \label{tab:4ferm}}
\end{table}

The Hermiticity of the physical Lagrangian enforces the coefficients of the operatorial class $\psi^2\vp^2 D$ to be real. Focusing on the one-loop contribution to the flavour-diagonal lepton-dipole momenta, one should also notice that the operatorial classes $(\bar LL)(\bar LL)$ and $(\bar RR)(\bar RR)$ could only give rise to real contributions. Consequently, they will not contribute (up to the one-loop level) to leptonic EDMs and will not be considered further.

Working in the broken phase rather than in the gauge basis, the two
operators of the $\psi^2 X\vp$ set can be rewritten using
\begin{align}
Q_{eB}&\rightarrow Q_{e\gamma}\hspace{0.1ex}c_W-Q_{eZ}\hspace{0.1ex} s_W,\\
Q_{eW}&\rightarrow -Q_{e\gamma}\hspace{0.1ex} s_W-Q_{eZ}\hspace{0.1ex} c_W,
\end{align}
where $s_W=\sin(\theta_W)$ and $c_W=\cos(\theta_W)$ are the sine and cosine of the weak mixing angle. 

The operator $Q_{e\vp}$ mixes with the dimension-four SM Lagrangian and redefines the relations among the Yukawa couplings and masses~\cite{Crivellin:2013hpa,Jenkins:2013zja,Jenkins:2013wua,Alonso:2013hga,Pruna:2014asa}:
\begin{align}\label{YukRed}
y_{l}\rightarrow 
\frac{\sqrt{2}m_l}{v}+\frac{v^2}{2\Lambda_{{\rm UV}}^2}C_{e\varphi}^{ll},
\end{align}
where both $y_{l}$ and $C_{e\varphi}^{ll}$ are complex parameters. For the correct gauge-invariant evaluation of the $Q_{e\vp}$ contribution to the leptonic EDMs (and flavour-diagonal dipole moments in general), the prescription of~\Eqn{YukRed} is crucial. 

The operator $Q_{le}$ requires additional consideration. The one-loop investigation of its chiral structure reveals a non-vanishing contribution in the naive anticommuting $\gamma_5$ scheme and, \emph{vice-versa}, a vanishing contribution when the Breitenlohner--Maison--'t Hooft--Veltman scheme is used\footnote{For a recent review of regularisation schemes see~\cite{Gnendiger:2017pys}, and for $\gamma_5$-related issues see~\cite{Gnendiger:2017rfh}.}. A scheme-independent result is only obtained when the two-loop anomalous dimensions are considered in the operatorial mixing, similarly as in~\cite{Ciuchini:1993fk,Crivellin:2017rmk}. Considering only the one-loop finite contribution, as was done in~\cite{Crivellin:2013hpa,Pruna:2014asa,Feruglio:2015yua,Pruna:2015jhf}, leads to non-physical results and should be avoided.

The natural energy scale for the measurement of the lepton EDM lies below the EW scale, which implies that $\mathcal{L}_{\rm SMEFT}$ should be matched to an effective Lagrangian invariant under QED and QCD symmetries that include higher-dimensional operators (see~\cite{Jenkins:2017jig} for the complete tree-level matching). For simplicity, only a single low-energy dipole operator will be considered in addition to the dimension-four QED and QCD Lagrangian\footnote{In principle, below the EW scale, one should also adopt a complete basis of dimension-six operators, as was done in~\cite{Crivellin:2017rmk}, for the study of muonic LFV processes. Again, such a complete analysis will be presented later in~\cite{future}.}, \emph{i.e.}
\begin{align}\label{lagem}
\mathcal{L}_{{\rm LEFT}} &= \left(2^{-3/4}G_F^{-1/2}\right)\frac{\mathcal{C}_{e\gamma}^{pr}}{\Lambda_{{\rm EW}}^2}
 (\bar l_p \sigma^{\mu\nu} e_r) F_{\mu\nu} \nonumber\\
&+\frac{\mathcal{C}_{S}^{prst}}{\Lambda_{{\rm EW}}^2}\left(\bar l_p P_R l_r\right) \left(\bar l_s P_R l_t\right)
+\frac{\mathcal{C}_{Tlu}^{prst}}{\Lambda_{{\rm EW}}^2}\left(\bar l_p \sigma^{\mu \nu} P_R l_r\right) \left(\bar u_s \sigma_{\mu \nu} P_R u_t\right) 
+\mbox{H.c.},
\end{align}
where $F_{\mu\nu}$ is the electromagnetic field-strength tensor and $\Lambda_{{\rm EW}}$ is the EW energy scale. 

In Section~\ref{sec:4}, the complete one-loop contribution of the dimension-six operators listed in Tables~\ref{tab:no4ferm} and~\ref{tab:4ferm} to the imaginary part of $\mathcal{C}_{e\gamma}^{ll}$ is presented. Several openly available tools were used to perform such a calculation in an automated way: the described model was implemented in {\tt FeynRules}~\cite{Alloul:2013bka} to obtain consistent Feynman rules\footnote{The model file is available upon request.}; the FeynArts interface of FeynRules was exploited to produce a model file for the {\tt FeynArts}~\cite{Hahn:2000kx} and {\tt FormCalc}~\cite{Hahn:1998yk,Hahn:2016ebn} packages; and the combined packages FeynArts/FormCalc were employed to generate non-integrated amplitudes, which are later elaborated with the symbolic manipulation system {\tt Form}~\cite{Kuipers:2012rf}.

\section{Electric-dipole moment of leptons}
\label{sec:3}
\noindent
This section defines the relevant observables and connects them with the effective coefficients as well as listing the current experimental bounds on lepton EDMs.

Following~\cite{Roberts:2010zz,Czarnecki:1900zz}, the structure of the fermionic dipole interaction can be defined as:
\begin{align}
\Gamma_\mu=F_1\left(q^2\right)\gamma_\mu
+iF_2\left(q^2\right)\sigma^{\mu\nu}q^\nu
-iF_3\left(q^2\right)\sigma^{\mu\nu}q^\nu\gamma_5+\left[\dots\right],
\end{align}
where the static charge and dipole moments are defined at $q^2\rightarrow 0$:
\begin{align}
F_1\left(0\right)&=Q_f e,\\
F_2\left(0\right)&=a_fQ_f \frac{e}{2m_f},\\
F_3\left(0\right)&=d_fQ_f.
\end{align}
In the previous equations, $a_f$ and $d_f$ are the anomalous magnetic and electric dipole moments of the fermion $f$, respectively.

At the lepton mass scale, the operator introduced in \Eqn{lagem} will contribute to the lepton dipole moments:
\begin{align}
a_l&= \frac{2}{e} \frac{2^{1/4} m_l }{ \sqrt{G_F} \Lambda_{{\rm EW}}^2}\Re\mathcal{C}_{e\gamma}^{ll},\\
d_l&= \frac{2^{1/4} }{\sqrt{G_F} \Lambda_{{\rm EW}}^2}\Im\mathcal{C}_{e\gamma}^{ll}.
\end{align}
Therefore, the imaginary part of the flavour-diagonal $\mathcal{C}_{e\gamma}$ coefficients represents the essential quantity to connect with the SMEFT dimension-six coefficients defined above the EW scale. Such quantities should be studied in the light of the current experimental limits for the lepton EDMs~\cite{Baron:2013eja,Bennett:2008dy,Inami:2002ah}:
\begin{align}
\left|d_e\right|&\le 8.7\cdot 10^{-29}\ e\hspace{0.2ex} {\rm cm}\simeq 2.1\cdot 10^{-16}\ {\rm GeV}^{-1}\left(90\%\ {\rm C.L.}\right),\\
\left|d_\mu\right|&\le 1.9\cdot 10^{-19}\ e\hspace{0.2ex} {\rm cm}\simeq 4.6\cdot 10^{-7}\ {\rm GeV}^{-1}\ \left(95\%\ {\rm C.L.}\right),\\
\left|d_\tau\right|&\le 4.5\cdot 10^{-17}\ e\hspace{0.2ex} {\rm cm}\simeq 1.1\cdot 10^{-4}\ {\rm GeV}^{-1}\ \left(95\%\ {\rm C.L.}\right).
\end{align}
These values can be trivially recast in terms of limits on the low-energy effective operator $\mathcal{C}_{e\gamma}$ defined at the lepton mass energy scale:
\begin{align}
\frac{\Im\mathcal{C}_{e\gamma}^{11}}{\Lambda_{{\rm EW}}^2}&<6.1\cdot 10^{-19}\ {\rm GeV}^{-2},\\
\frac{\Im\mathcal{C}_{e\gamma}^{22}}{\Lambda_{{\rm EW}}^2}&<1.3\cdot 10^{-9}\ {\rm GeV}^{-2},\\
\frac{\Im\mathcal{C}_{e\gamma}^{33}}{\Lambda_{{\rm EW}}^2}&<3.1\cdot 10^{-7}\ {\rm GeV}^{-2}.
\end{align}
The equations above can not be directly translated into the SMEFT picture with a naive tree-level approach. Instead, the strategy of regions~\cite{Beneke:1997zp} and standard renormalisation-group evolution (RGE) technologies (both embedded in a consistent perturbative approach) should be exploited to connect these constraints to SMEFT dimension-six coefficients defined at higher energies. The next section outlines the main ideas behind such a procedure.

\section{Anomalous dimensions and matching between LEFT and SMEFT}
\label{sec:4}
\noindent
In this section, the items required to connect the effective coefficients at different energy scales are described.

To identify the leading contribution to the operatorial mixing into dipole operators, it is necessary to evaluate their RG equations. For $\mathcal{L}_{{\rm LEFT}}$, the result was presented in~\cite{Crivellin:2016ebg,Crivellin:2017rmk} and reads:
\begin{align}
16\pi^2\left(2^{-3/4}G_F^{-1/2}\right)\dot{\mathcal{C}}_{e\gamma}^{ll}&=
\left(10 Q_l^2+\sum_f\frac{4}{3}N_c Q_f^2\right) e^2
\left(2^{-3/4}G_F^{-1/2}\right)\mathcal{C}_{e\gamma}^{ll}\nonumber \\
&-2Q_le\sum_{r}m_r\mathcal{C}_{S}^{rllr}
+16 e \sum_{c} m_{c} \mathcal{C}_{Tlu}^{llcc},
\end{align}
where the indices $l$ and $r$ indicate a sum over the leptonic flavours and $c$ indicates a sum over the $u$-type quarks (with the top-quark integrated out). For $\mathcal{L}_{{\rm SMEFT}}$, the result was presented in~\cite{Jenkins:2013zja,Jenkins:2013wua,Alonso:2013hga,Pruna:2014asa}, and if only the gauge couplings and the top-quark Yukawa coupling are kept, then it reads:
\begin{align}\label{CEG_RGE}
&\quad
16\pi^2\dot{C_{e\gamma}^{ll}}
\simeq \left(\frac{47 e^2}{3}+\frac{e^2}{4 c_W^2}
-\frac{9 e^2}{4 s_W^2}+3 y_t^2\right)C_{e\gamma}^{ll}
+6e^2\left(\frac{c_W }{s_W}-\frac{s_W}{c_W}\right)C_{eZ}^{ll}
+16 e \sum_{c}y_cC^{(3)}_{ll cc}.
\end{align}

The tree-level and one-loop matching of the relevant operators at the EW energy scale are described in Tables~\ref{tab:tlmatching} and~\ref{tab:olmatching}, respectively. The latter is obtained by straightforward application of the strategy of regions~\cite{Smirnov:1994tg,Beneke:1997zp,Smirnov:2002pj}.

\begin{table}[!ht] 
\centering
\renewcommand{\arraystretch}{1.2}
\btb{||c|c||} 
\hline 
\hline 
Coefficient &  Tree-level matching at the EW scale\\
\hline
\ & 
\ 
\\[-2ex]
$\Im C_{e\gamma}^{ll}$ & 
$
\Im{C_{e\gamma}^{ll}}
$
\\[2ex]
$\Im{\mathcal{C}_{S}^{rllr}}$ & 
$
\begin{aligned}
-\frac{v^2}{4m_H^2}\left(y_{lr}\Im{C_{e\varphi}^{rl}}\delta_{lr}
+y_{rl}\Im{C_{e\varphi}^{lr}}\delta_{rl}\right)
\end{aligned}
$
\\[2ex]
$\Im{\mathcal{C}_{Tlu}^{llcc}}$ & 
$
-\Im{C_{lequ(3)}^{llcc}}
$
\\[2ex]
\hline
\hline
\etb
\caption{Tree-level matching at the EW energy scale between the coefficients of $\mathcal{L}_{{\rm SMEFT}}$ and $\mathcal{L}_{{\rm LEFT}}$. \label{tab:tlmatching}}
\end{table}

\begin{table}[!ht] 
\centering
\renewcommand{\arraystretch}{1.2}
\btb{||c|c||} 
\hline 
\hline 
Coefficient & Hard one-loop contribution to $\Im{\mathcal{C}_{e\gamma}^{ll}}$ \\
\hline
\ & 
\ 
\\[-2ex]

$C_{e\varphi}^{ll}$ & 
$
\begin{aligned}
\frac{e y_l^2}{64  \pi^2 }\frac{v^2}{m_H^2}
\left(-3 +4 \log\left[\frac{m_H}{\Lambda_{\rm{EW}}}\right]\right)\Im{C_{e\varphi}^{ll}}
\end{aligned}
$
\\[2ex]
$C_{eZ}^{ll}$
& 
$
\begin{aligned}
\frac{\alpha}{8\pi c_W  s_W}  \left(-3 c_W^2+3 s_W^2+8 c_W^2 \log\left[\frac{m_W}{\Lambda_{\rm{EW}}}\right]+4 \left(c_W^2-3 s_W^2\right) \log\left[\frac{m_Z}{\Lambda_{\rm{EW}}}\right]\right)\Im{C_{eZ}^{ll}}
\end{aligned}
$
\\[2ex]
$C_{lequ(3)}^{llcc}$
& 
$
\begin{aligned}
-\frac{e y_c }{\pi ^2}\log\left[\frac{m_c}{\Lambda_{\rm{EW}}}\right]\Im{C_{lequ(3)}^{llcc}}
\end{aligned}
$

\\[2ex]

\hline
\hline
\etb
\caption{One-loop matching at the EW energy scale between the coefficients of $\mathcal{L}_{{\rm SMEFT}}$ and $\Im \mathcal{C}_{e\gamma}^{ll}$. \label{tab:olmatching}}
\end{table}

These elements are pieced together in the next section in a consistent phenomenological overview of experimental limits acting on the effective coefficients defined at higher energy scales.

\section{Results}
\label{sec:5}
\noindent
The main running and mixing effects give rise to the set of uncorrelated limits in Table~\ref{tab:res} evaluated at the EW scale $\Lambda_{{\rm EW}}\simeq 100$ GeV. Any BSM theory matching with the SMEFT coefficients involved in this analysis should respect these limits at the EW scale.
\begin{table}[!ht] 
\centering
\renewcommand{\arraystretch}{1.2}
\btb{||c|c||c|c||c|c||} 
\hline 
\hline 
Coefficient & Limits from $d_e$ & Coefficient & Limits from $d_\mu$ & Coefficient & Limits from $d_\tau$ \\
\hline
\ &\ &\ & \ & \ & \ 
\\[-3ex]
$\Im C_{e\gamma}^{11}$ & 
$
\begin{aligned}
6.1\times 10^{-15}
\end{aligned}
$
&
$\Im C_{e\gamma}^{22}$
&
$
\begin{aligned}
1.3\times 10^{-5}
\end{aligned}
$
&
$\Im C_{e\gamma}^{33}$
&
$
\begin{aligned}
3.1\times 10^{-3}
\end{aligned}
$
 \\[2ex]

$\Im C_{e\varphi}^{11}$ & 
$
\begin{aligned}
8.2\times 10^{-3}
\end{aligned}
$
&
$\Im C_{e\varphi}^{22}$
&
N/A
&
$\Im C_{e\varphi}^{33}$
&
N/A
\\[2ex]
$\Im C_{eZ}^{11}$ & 
$
\begin{aligned}
5.1\times 10^{-12}
\end{aligned}
$
&
$\Im C_{eZ}^{22}$
&
$
\begin{aligned}
1.1\times 10^{-2}
\end{aligned}
$
&
$\Im C_{eZ}^{33}$
&
N/A
\\[2ex]
$\Im C_{lequ(3)}^{1122}$ & 
$
\begin{aligned}
6.0\times 10^{-12}
\end{aligned}
$
&
$\Im C_{lequ(3)}^{2222}$
&
$
\begin{aligned}
1.3\times 10^{-2}
\end{aligned}
$
&
$\Im C_{lequ(3)}^{3322}$
&
N/A
\\[2ex]
$\Im C_{lequ(3)}^{1111}$ & 
$
\begin{aligned}
1.3\times 10^{-9}
\end{aligned}
$
&
$\Im C_{lequ(3)}^{2211}$
&
N/A
&
$\Im C_{lequ(3)}^{3311}$
&
N/A
\\[2ex]
 \hline
\hline
\etb
\caption{Limits on SMEFT effective coefficients defined at the EW energy scale from leptonic EDMs. \label{tab:res}}
\end{table}

Interestingly, some of these constraints are orders of magnitude more stringent than the analogous ones extracted using tree-level collider analysis. For example, the limits on $C_{lequ(3)}^{1111}$ presented here and in~\cite{Falkowski:2017pss} can be compared. This enforces the conclusion that adopting a perturbative approach with higher-order leading contributions to precision observables can give radically different qualitative and quantitative results with respect to a tree-level collider analysis.
\begin{figure}[!ht]
\begin{center}
\includegraphics[width=0.48\textwidth]{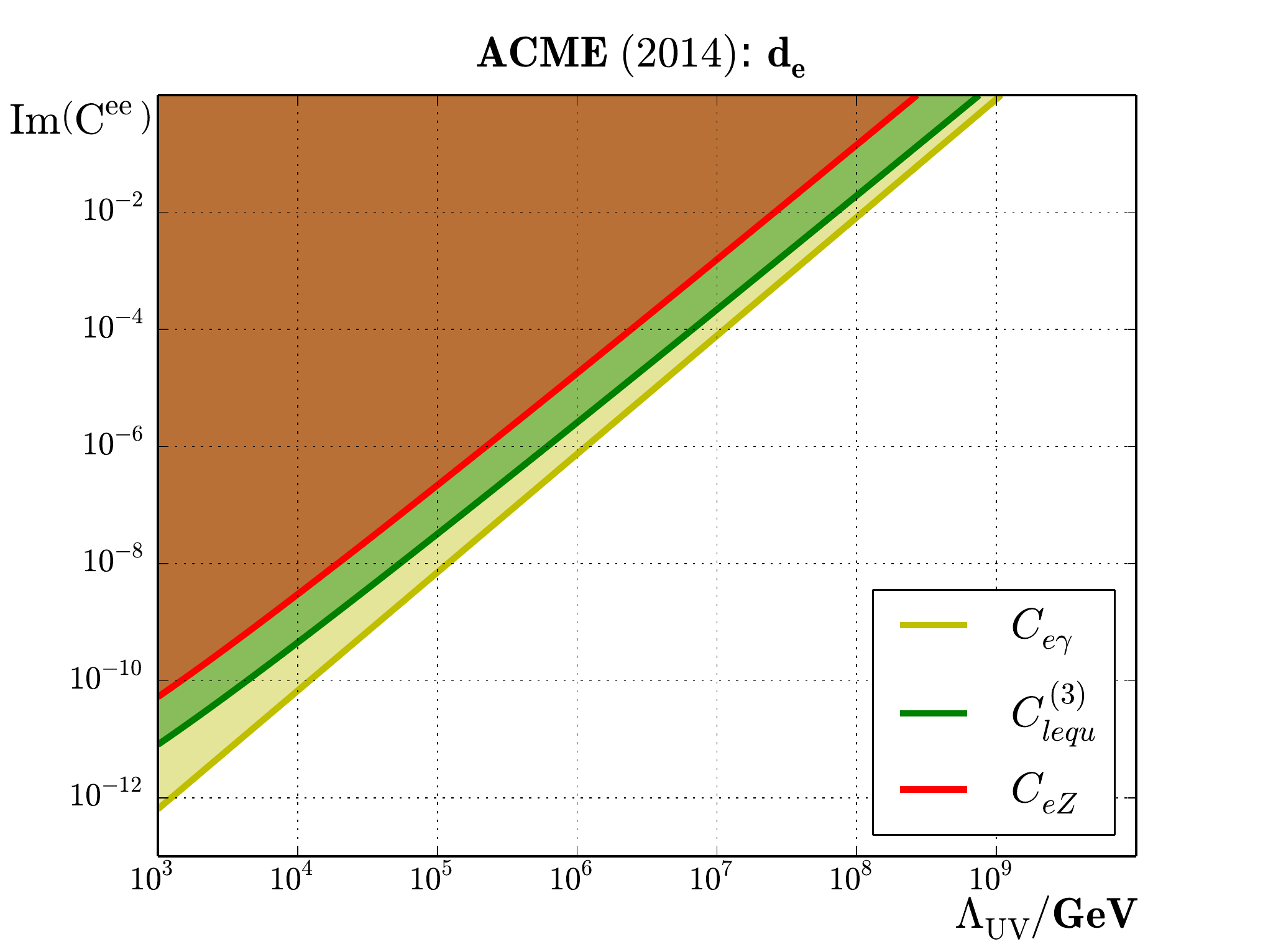}
\includegraphics[width=0.48\textwidth]{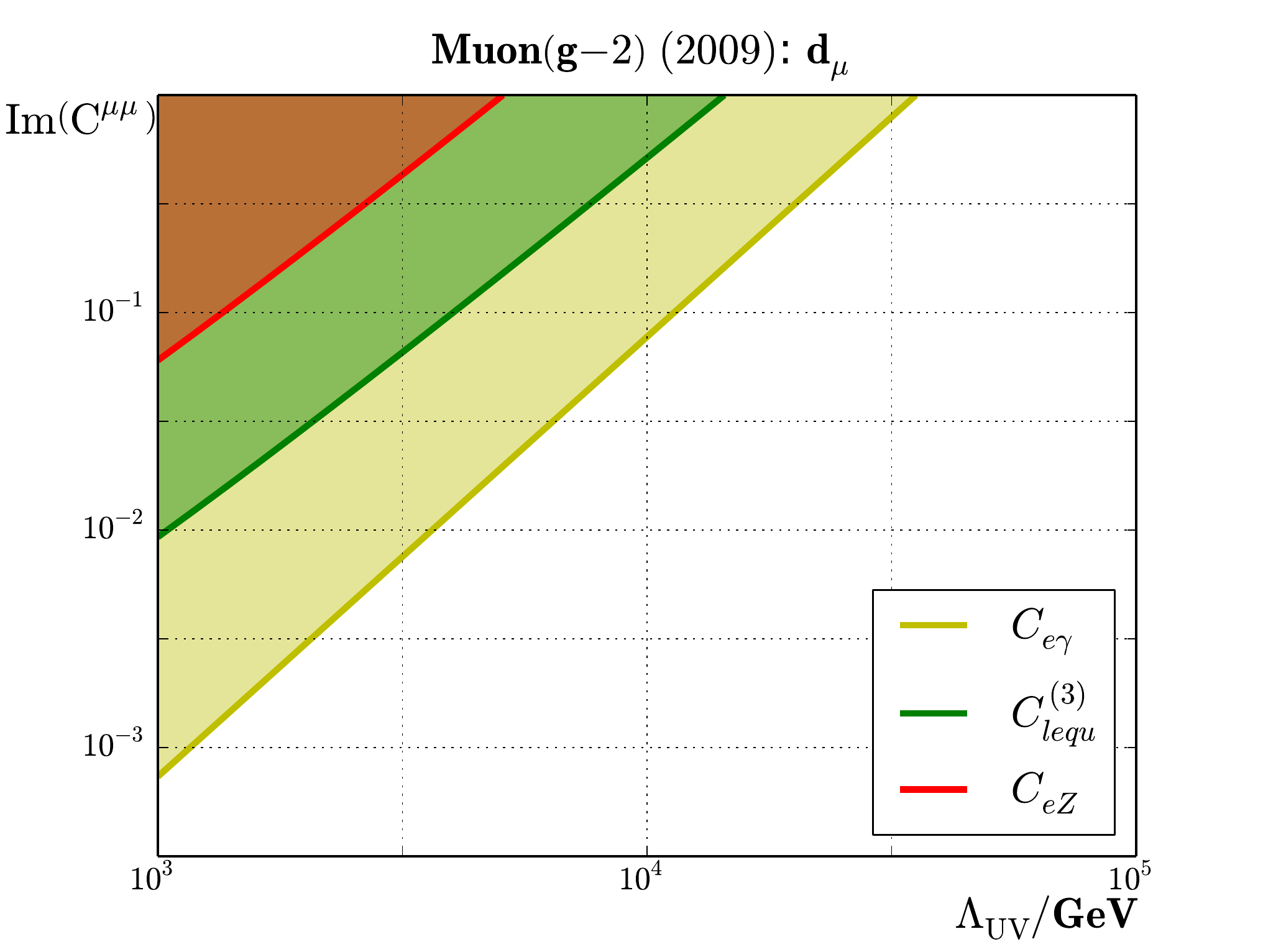}
\end{center}
\caption{Limits on SMEFT effective coefficients defined at the $\Lambda_{\rm UV}$ energy scale from leptonic EDMs.}
\label{fig:1}
\end{figure}

In Figure~\ref{fig:1}, the limits on the effective contribution at the EW scale to the imaginary part of the flavour-diagonal $C_{e\gamma}$ coefficients from the ACME (left panel) and Muon $g-2$ (right panel) collaborations are plotted. The leading one-loop logarithms give constraints on $C_{e\gamma}$, $C_{eZ}$ and $C_{lequ(3)}^{ll33}$ defined at higher energies $\Lambda_{{\rm UV}}$.

\section{Conclusion}
\label{sec:6}
\noindent
CP violation in the charged lepton sector of the SM was studied in a SMEFT framework with the inclusion of dimension-six effective operators.

The current experimental limits on lepton EDMs were translated into constraints on the imaginary part of the complex effective coefficients defined at the EW scale and above. Exploiting a systematic one-loop SMEFT analysis, new bounds on the coefficients above were presented. 

The most stringent bounds are obtained in connection with the electron EDM due to the extreme accuracy on the experimental limit, while the current determination of the $\tau$ EDM does not allow for reasonable bounds. This could potentially improve in the future with the advent of innovative ideas for a better determination of the $\tau$ EDM~\cite{Eidelman:2016aih}.

However, this study is far from being comprehensive by regarding all the possible features of a complete SMEFT analysis. Important multi-loop effects are neglected, including the two-loop mixing of $Q_{le}$ into the dipole operator $Q_{e\gamma}$. Furthermore, non-trivial mixing and matching effects can arise through consideration of the complete set of dimension-six SMEFT operators rather than the leptonic subset analysed here. These gaps in the analysis will be explored in a future publication~\cite{future}.
\section*{Acknowledgements}
\noindent
This research was supported by the SNSF under contract 200021\_160156. I am profoundly grateful to Matteo Fael and Adrian Signer for having shared useful comments on this study, thereby giving me the opportunity to improve this manuscript.

\bibliographystyle{JHEP}
\bibliography{Pruna_G_M}

\end{document}